\documentclass[acmsmall, authorversion=true]{acmart}
\AtBeginDocument{%
  }

\setcopyright{acmlicensed}
\copyrightyear{2026}
\acmYear{2026}
\acmDOI{}

\acmJournal{TAIS}

\usepackage[utf8]{inputenc} %
\usepackage[T1]{fontenc}    %
\usepackage{booktabs}       %
\usepackage[singlelinecheck=true]{subcaption}

\usepackage{amsfonts, amsmath, amssymb}       %
\usepackage{nicefrac}       %
\usepackage{microtype}      %
\usepackage{lipsum}
\usepackage{fancyhdr}       %
\usepackage{graphicx}       %
\usepackage{multirow}
\usepackage{float}
\usepackage{wrapfig}
\usepackage{todonotes}
\usepackage{comment}
\usepackage{tikz}
\usepackage{threeparttablex}

\usetikzlibrary{positioning, fit, arrows.meta, backgrounds}

\newcommand{\rv}[1]{#1}

\graphicspath{{media/}}     %

\newcommand{\dd}{D^{\dagger}D}

\title{Matrix-free Neural Preconditioner for the Dirac Operator in Lattice Gauge Theory}
\titlenote{Preprint No. MIT-CTP/5915, INT-PUB-25-022}

\author{Yixuan Sun }
\email{yixuan.sun@anl.gov}
\orcid{0000-0003-1109-3380}
\author{Srinivas Eswar}
\email{seswar@anl.gov}
\orcid{0000-0002-3418-7796}
\affiliation{%
  \institution{Argonne National Laboratory}
  \city{Lemont}
  \state{Illinois}
  \country{USA}
}

\author{Yin Lin}
\email{yin01@mit.edu}
\affiliation{%
  \institution{Massachusetts Institute of Technology}
  \city{Lemont}
  \state{Illinois}
  \country{USA}
}

\author{William Detmold}
\email{wdetmold@mit.edu}
\orcid{0000-0002-0400-8363}
\author{Phiala Shanahan}
\email{pshana@mit.edu}
\orcid{0000-0002-0916-7603}
\affiliation{%
  \institution{Massachusetts Institute of Technology \& The NSF AI Institute for Artificial Intelligence and Fundamental Interactions}
  \city{Cambridge}
  \state{Massachusetts}
  \country{USA}
}

\author{Yang Liu}
\email{liuyangzhuan@lbl.gov}
\orcid{0000-0003-3750-1178}
\author{Xiaoye Li}
\email{xsli@lbl.gov}
\orcid{0000-0002-0747-698X}
\affiliation{%
  \institution{Lawrance Berkeley National Laboratory}
  \city{Berkeley}
  \state{California}
  \country{USA}
}

\author{Prasanna Balaprakash}
\email{pbalapra@ornl.gov}
\orcid{0000-0002-0292-5715}
\affiliation{%
  \institution{Oak Ridge National Laboratory}
  \city{Oak Ridge}
  \state{Tennessee}
  \country{USA}
}

\renewcommand{\shortauthors}{Sun et al.}

\begin{abstract}

Linear systems arise in generating samples and in calculating observables in lattice quantum chromodynamics~(QCD). Solving the Hermitian positive definite systems, which are sparse but ill-conditioned, involves using iterative methods, such as Conjugate Gradient (CG), which are time-consuming and computationally expensive. Preconditioners can effectively accelerate this process, with the state-of-the-art being multigrid preconditioners. However, constructing useful preconditioners can be challenging, adding additional computational overhead, especially in large linear systems. We propose a framework, leveraging operator learning techniques, to construct linear maps as effective preconditioners. The method in this work does \emph{not} rely on explicit matrices from either the original linear systems or the produced preconditioners, allowing efficient model training and application in the CG solver. In the context of the Schwinger model (\(U(1)\) gauge theory in 1+1 spacetime dimensions with two degenerate-mass fermions), this preconditioning scheme effectively decreases the condition number of the linear systems  and approximately halves the number of iterations required for convergence in relevant parameter ranges. We further demonstrate the framework learns a general mapping dependent on the lattice structure which leads to zero-shot learning ability for the Dirac operators constructed from gauge field configurations of different sizes.
\end{abstract}



\begin{document}
\maketitle
\fancyfoot[RO,LE]{\footnotesize Author's accepted version. ACM Trans. AI Sci., to appear.}

\begin{tikzpicture}[remember picture,overlay]
  \node[anchor=north east,xshift=-2cm, yshift=-1cm] at (current page.north east) {%
    MIT-CTP/5915, INT-PUB-25-022
  };
\end{tikzpicture}

\section{Introduction}

Lattice quantum field theory (LQFT) provides a non-perturbative framework for
studying quantum field theories by discretizing spacetime on a finite lattice
as an intermediate step~\cite{usqcd2019hadrons}. This approach makes it
possible to investigate strongly coupled field theories numerically, including
in particular quantum chromodynamics (QCD), the theory of the strong
interaction in particle and nuclear physics. Lattice QCD (LQCD) has been
instrumental in computing hadronic properties from the first principles,
contributing to precise predictions of many observables within the Standard
Model~\cite{particle2020review, usqcd2019hot}.

In LQFT calculations involving fermions, a central computational task is the
repeated solution of large linear systems of the form

\begin{equation}
  \label{eqn:lin}
  A x = b,
\end{equation}
where the matrix $A = \dd{[U_\mu(x), \kappa]}$ encodes the discretized Dirac
normal operator on a given gauge field background $U_\mu(x)$, \rv{ $\kappa$ is
the hopping parameter, $D$ is the Dirac operator, $D^\dagger$ is its conjugate
transpose, and $[\cdot]$ represents the functional dependence.} These matrices
are typically sparse, complex, and very large, with dimensions that can exceed
$10^8$ depending on the lattice size and fermion formulation.
In practice, in a given calculation, hundreds or thousands of \(A\)'s will be
constructed, with the same dimensionality and sparsity, and they are often
ill-conditioned. In addition, (\ref{eqn:lin}) must typically be solved for
hundreds of right-hand sides for each \(A\). As a result, the total number of
linear solves in a typical calculation can reach into the millions. These
solves account for a significant portion of the overall computational
cost~\cite{joo2019status}. Improving the efficiency of these solves remains a
key priority for advancing the reach of LQFT, and motivates ongoing efforts in
algorithm development, preconditioning strategies, and the use of
high-performance computing resources.

Solving such linear systems often relies on iterative methods, such as the
Conjugate Gradient~(CG) algorithm~\cite{saad2003iterative}, which iteratively
produces approximated solutions until target accuracy is reached. However, the
number of iterations required is determined by the condition number of the
linear system, and efficiently solving the ill-conditioned systems that arise
in LQFT, therefore, still remains a challenge. This is particularly evident in
the context of lattice QCD calculations for small physical lattice spacings and
light quark masses~\cite{cali_neural-network_2023}. Accelerating the iterative
solvers prompts the development of effective preconditioning techniques such as
incomplete LU and algebraic multigrid (AMG)
methods~\cite{brannick_bootstrap_2014, brannick2008adaptive,
  babich2010adaptive,Brower:2018ymy,Brower:2020xmc,Babich:2011np,Clark:2016rdz,Frommer:2013fsa,Boyle:2021wcf,Alexandrou:2016izb,Brannick:2014vda,Brower:1991xv,Hulsebos:1990er,Whyte:2025sjw,Gruber:2024cos,Boyle:2024pio,Boyle:2024nlh,Frommer:2022fwk,Richtmann:2022fwb,Richtmann:2019eyj,Espinoza-Valverde:2022pci}.

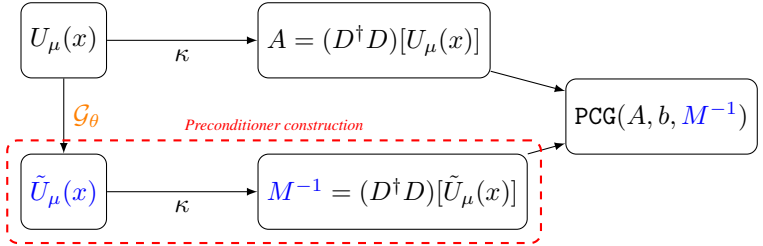
\begin{figure}[!htbp]
  \centering
  \begin{tikzpicture}[
    data/.style={ rectangle, draw, rounded corners, minimum size=1cm, fill=white,
        line width=0.8pt },
    operator/.style={ rectangle, draw, rounded corners, minimum size=1cm, line
        width=1.2pt, densely dashed, fill=orange!5 },
    solver/.style={ rectangle, draw, rounded corners, minimum size=1cm, line
        width=0.8pt, fill=white },
    arr/.style={-{Latex[length=2.5mm]}, thick} ]

    \node[data] (input) at (0, 0) {$U_{\mu}(x)$};
    \node[data, below=1.2cm of input] (new_input) {$\textcolor{blue}{\tilde{U}_\mu(x)}$};

    \node[operator, right=2.2cm of input] (system) {$A = (\dd)[U_{\mu}(x), \kappa]$};
    \node[operator, right=2.2cm of new_input] (precond) {$\textcolor{blue}{\Gamma} = (\dd){[\tilde{U}_\mu(x), \kappa]}$};

    \node[solver, right=0.6cm of precond, yshift=1.1cm] (preconditioned) {$\texttt{PCG}(A, b, \textcolor{blue}{\Gamma})$};

    \path[arr] (input) edge node[right] {$\textcolor{orange}{\mathcal{G}_{\theta}}$} (new_input);
    \path[arr] (input) edge (system);
    \path[arr] (system) edge (preconditioned);
    \path[arr] (new_input) edge (precond);
    \path[arr] (precond) edge (preconditioned);

    \node[draw, rectangle, rounded corners, dashed,
    color=red, thick,
    label=above:{\small \it \color{red} Preconditioner Construction},
    fit=(new_input) (precond), inner sep=6pt] {};

    \node[anchor=north west, font=\footnotesize] (legend_title) at ([yshift=-1.0cm, xshift=-0.3cm]new_input.south west) {};

    \node[data, minimum size=0.5cm, right=0.0cm of legend_title, anchor=west] (leg_data) {};
    \node[right=0.15cm of leg_data, font=\footnotesize, anchor=west] (leg_data_label) {Field Configuration};

    \node[operator, minimum size=0.5cm, right=0.6cm of leg_data_label, anchor=west] (leg_op) {};
    \node[right=0.15cm of leg_op, font=\footnotesize, anchor=west] (leg_op_label) {Linear Operator (mat-vec product)};

  \end{tikzpicture}
  \caption{Matrix-free neural preconditioner for the discretized Dirac operator
    calculated from a given gauge configuration $U_{\mu}(x)$. The trained neural network $\mathcal{G}_\theta$ outputs another set of
    configurations $\tilde{U}_\mu(x)$ and uses the same discretization scheme \rv{(same $\kappa$)} for the
    original system to construct the preconditioning operator $\Gamma$. For a given linear system, we can directly pass
    the original and preconditioning operators, $A$ and $\Gamma$, as well as the right hand side $b$ to a Preconditioned Conjugate Gradient (PCG) solver. All
    processes involved do not use explicit matrices.
  }
  \label{fig:overview}
\end{figure}

Preconditioners improve the spectral properties of the matrix under
consideration, aiming to reduce the number of iterations required for
convergence. However, constructing an effective preconditioner typically
requires a deep understanding of the underlying structure of the linear
equations and often requires extra computational overhead, as a tailored
preconditioner must be built or updated for each specific system. This trade-off
between achieving faster convergence and managing additional setup cost is a
well-known challenge in numerical linear algebra~\cite{saad2003iterative,
barrett1994templates}. In recent years, machine learning (ML) based
preconditioners have emerged~\cite{hausner2024neural, yusuf_constructing_2024,
  cali_neural-network_2023, li_neuralpcg_2022, azulay_multigrid-augmented_2022,
  sappl_deep_2019, tang_graph_nodate} to address this issue. These methods learn
from existing data in both supervised and unsupervised ways and are able to
generalize to other systems in the same problem families. However, these methods
generally still rely on explicit matrices for both the linear systems and their
preconditioners, making it memory and computationally infeasible for large
systems, such as those found in LQFT applications. \rv{While there exists an
ML-based preconditioner learning framework that does not need full explicit
matrices \cite{chen2025graph}, the original operator is still required for
both training and preconditioner construction}. In this work,  we propose an
ML-based preconditioner framework \rv{that operates entirely in the space of
gauge field configurations}, shown in Figure~\ref{fig:overview}.  We adopt an
operator learning approach to construct the preconditioners \rv{solely from the
input gauge configuration, directly applicable as matrix-vector products without
assembling or storing explicit matrices.} While this framework is general, in
this work, we focus on the Schwinger model \(U(1)\) gauge theory in 1+1
dimensions with two flavors of mass-degenerate fermions as a proof of concept to
show the viability of the proposed technique. We demonstrate the performance of
the proposed method by solving the Wilson-Dirac normal equations with different
lattice sizes and show the resulting preconditioners effectively reduce the
number of iterations required for convergence in the CG solver. Moreover, models
trained on a single lattice gauge configuration ensemble can immediately act as
effective preconditioners for systems of other lattice sizes, achieving
zero-shot transfer. This eliminates the need for retraining for unseen gauge
configurations, at least within the parameter ranges explored in this work. We
summarize the contribution of this work as follows

\begin{itemize}
  \item We propose a framework leveraging an \emph{operator learning} method to
        construct effective preconditioners for solving Wilson-Dirac normal equations.
  \item The framework is completely \emph{matrix-free} where it only operates on
        lattice gauge field configurations and \emph{does not} construct explicit
        matrices for either the linear operators or their preconditioners.
  \item We train the model in an unsupervised way and use random projections to
        formulate a loss function that is effective and efficient in model training,
        avoiding expensive condition number computation.
  \item The trained models produce preconditioners that can be directly integrated into
        CG solvers, approximately halving the required number of iterations for
        convergence.
  \item The proposed framework learns a general mapping applicable for \emph{unseen}
        lattice ensembles of different parameters and sizes, attaining zero-shot
        performance comparable to that of networks trained for specific action
        parameters.
\end{itemize}

In the remainder of this paper, we briefly discuss the related concepts in
preconditioning, machine learning-based preconditioners, and operator learning
frameworks in Section~\ref{sec:related-work}. We then describe the details of
the proposed approach in Section~\ref{sec:methods} and present and analyze its
performance on solving Dirac equations for different lattice geometries and
parameters in Section~\ref{sec:exp}. Finally, we summarize this work and
discuss future directions in Section~\ref{sec:conclusion}.

\section{Related work}\label{sec:related-work}

\paragraph{General preconditioners}

The essence of preconditioning is converting a system of linear equations which
is not readily solvable into one which is easier or faster to solve
\cite{wathen2015preconditioning}. Common preconditioning techniques are
diagonal scaling and incomplete factorizations (e.g., incomplete LU) of the
input coefficient matrix \cite{saad2003iterative}. Since these techniques are
often used in conjunction with iterative Krylov space methods, the choice of
the iterative method impose different properties on the preconditioner
\cite{greenbaum1997iterative,van2003iterative,saad2003iterative,olshanskii2014iterative}.
For the normal equations based conjugate gradient method, as in this work,
popular preconditioners include the incomplete Cholesky, incomplete shifted
Cholesky, and the incomplete LQ factorizations \cite{saad2003iterative}. A
special note must be made for preconditioners arising from discretizations of
partial differential equations (PDEs), where preconditioners are formed on the
basis of the underlying operators
\cite{mardal2011preconditioning,gunnel2014note, malek2014preconditioning}.
Multigrid preconditioners are the natural and most commonly used techniques in
this setting \cite{wathen2015preconditioning}. For LQCD, multigrid
preconditioners are the dominant approach in current state-of-the-art
calculations~\cite{brannick_bootstrap_2014, brannick2008adaptive,
  babich2010adaptive,Brower:2018ymy,Brower:2020xmc,Babich:2011np,Clark:2016rdz,Frommer:2013fsa,Boyle:2021wcf,Alexandrou:2016izb,Brannick:2014vda,Brower:1991xv,Hulsebos:1990er,Whyte:2025sjw,Gruber:2024cos,Boyle:2024pio,Boyle:2024nlh,Frommer:2022fwk,Richtmann:2022fwb,Richtmann:2019eyj,Espinoza-Valverde:2022pci}.
For more comprehensive reviews of preconditioners and their applications, we
direct the interested reader to
\cite{benzi2002preconditioning,saad2003iterative,wathen2015preconditioning}.

\paragraph{ML-based preconditioners}

As constructing effective preconditioners requires domain expertise and
repetition for individual linear systems, ML-based methods have shown
flexibility by directly learning from data, and domain adaptability by
generalizing to systems for the same problem family. While some work, inspired
by the classical preconditioning approaches, has been focusing on using neural
networks to perform matrix factorization as
preconditioners~\cite{yusuf_constructing_2024, hausner_neural_2024} or to
replace a multigrid cycle~\cite{azulay_multigrid-augmented_2022}, most use the
inverse approximating property of preconditioners to train domain-specific or
general-purpose neural network (NN)-based
preconditioners~\cite{chen2025graph, ackmann_machine-learned_2020,
  sappl_deep_2019, li_neuralpcg_2022}. \rv{Another line of work combines machine learning and domain decomposition to construct preconditioners. See~\cite{klawonn2024machine} for a recent survey. More broadly, neural operators have been used to learn preconditioners for Krylov solvers~\cite{kopanivcakova2025deeponet, Rudikov2024-dr, Kopanicakova2025-du}.}
In the space of lattice quantum field theory,
\cite{lehner_gauge-equivariant_2023, lehner2024gauge} leverages gauge equivariance neural
networks to construct linear maps as effective multigrid preconditioners for
Dirac equations. However, it requires retraining on unseen gauge configurations
within a given gauge ensemble. To exploit the nonlinear function approximation
ability of neural networks, \cite{cali_neural-network_2023} utilizes 4D
convolutional networks to directly transform the Wilson-Dirac normal matrices
to the corresponding preconditioners. They also demonstrate the volume transfer
capability for gauge configurations with different lattice sizes. Our work
closely follows~\cite{cali_neural-network_2023}, but with a key difference:
instead of operating on explicit matrices, our framework is
matrix-free and more amenable to solving large problems, as necessary for
state-of-the-art LQCD calculations. In addition, this approach produces
preconditioning operators that are immediately applicable in iterative solvers
without additional linear solves while retaining the volume transfer
capability.

\paragraph{Operator learning}

In scientific domains, where the transformations are often between
infinite-dimensional function spaces, they usually are not able to capture the
underlying function spaces sufficiently and are limited to the structures of
available data~\cite{cybenko_approximation_1989, hornik_multilayer_1989}. As a
result, such models have subpar generalization and require retraining or
substantial fine-tuning for data with a different resolution. Operator learning
frameworks~\cite{li_fourier_2021, lu_learning_2021, kovachki_neural_2023}
address this issue by learning the mapping between function spaces. Because of
their ability to exploit the function structures and capture both local and
global behaviors, operator learning models have shown success in various
domains~\cite{patel_physics-informed_2021, goswami_physics-informed_2021,
  sun_deepgraphonet_2023}. \rv{One notable property of operator learning
  frameworks is that they are resolution-independent in their formulation,
  invariant to discretization as they learn the mapping between functions. Thus,
  they can be queried at arbitrary resolutions without retraining.} This is
achieved generally by learning integrating kernels that perform transformations
between the function spaces~\cite{boulle_mathematical_2023}, which naturally is
applicable to different discretizations. Our work adopts the Fourier Neural
Operator~\cite{li_fourier_2021} and Fully Convolutional
Network~(FCN)~\cite{long_fully_2015} to capture the mapping between the space
of gauge configurations to a related embedding space as part of the proposed
framework to construct effective preconditioners for the Wilson-Dirac normal
equations. This is the key component enabling the successful volume transfer to
lattice gauge field configurations with various lattice geometries.

\section{Methods}\label{sec:methods}

In this section, we formulate the learning task as an operator learning
problem, employing FNO and FCN architectures to model the mapping between input
and output configurations shown in Figure~\ref{fig:overview}. We further
introduce an efficient unsupervised loss function designed to train the
networks such that the linear operator derived from the output configuration
approximates the inverse of that associated with the input configuration.

\subsection{Problem formulation}
\label{ssec:probform}

The lattice discretization of the two-flavor Schwinger model used in this study
is defined by the standard plaquette gauge action and the Wilson fermion action
corresponding to two degenerate fermions:
\begin{align}
  \begin{split}
    S = & -\beta\sum_{x\in \Lambda_L}\text{Re}\big(P(x)\big) +
    \sum_{f=0}^1\sum_{x,y\in\Lambda_L} \overline{\psi}_{f}(x) D_{x,y}\psi_{f}(y),
  \end{split}
\end{align}
where the $d=2$ spacetime lattice of finite extent $L$ in each direction is
given by \(\Lambda_L = \{x=an| n \in \mathbb{Z}_L^d\}\) for lattice spacing
$a$, $U_{\mu}(x) \in \text{U}(1)$ is the complex gauge field where ${\mu\in\{1,
      2\}}$ labels the spatial and temporal components and $\psi_{f}(x)$,
$\overline{\psi}_{f}(x)$ are two-component Wilson fermion fields with flavor
indices $f\in\{0,1\}$. The plaquette appearing in the gauge action is
\begin{equation}
  P(x) =  U_{1}(x)U_{2}(x+\hat{1})U^*_{1}(x+\hat{2})U^*_{2}(x),
\end{equation}
\noindent where $\hat{j}$ is a unit vector in the $j$ direction,
and the Wilson discretization of the Dirac operator~\cite{wilson1974confinement} is

\begin{equation}
  \label{eqn:dirac}
  \begin{aligned}
    D_{x,y} & = (m+2r)\delta_{x,y}
    -
    \frac{1}{2}\sum_{\mu=1}^2\bigg(
    (1-\gamma_\mu) U_{\mu}(x)\delta_{x+\hat{\mu},y}
    +
    (1+\gamma_{\mu})U^*_{\mu}(x-\hat{\mu})
    \delta_{x-\hat{\mu},y}
    \bigg),
  \end{aligned}
\end{equation}
where $\gamma_1$ and $\gamma_2$ are Euclidean gamma matrices in two dimensions. The specific representation  used in this study is provided by the Pauli matrices: $\gamma_1 = \sigma_1$ and $\gamma_2 = \sigma_2$ with  $\gamma_5 = i\gamma_1\gamma_2 = -\sigma_3$.
Periodic spatial boundary conditions are used for all fields and the fermion(boson) fields are anti-periodic(periodic) in the temporal direction.

The Wilson-Dirac operator depends on the bare fermion mass $m$ and Wilson term
$r$ and implicitly on the bare gauge coupling $\beta$. In this work, $r=1$
throughout and $m$ is implemented through $\kappa=(2(m+2))^{-1}$.
Since \(D\) itself is not Hermitian for the Wilson fermion discretization, we
combine it with its conjugate transpose to produce a Hermitian system $\dd x =
  D^{\dagger}b$ with the solution \(x\) to the original system ($Dx=b$) easily
constructed thereafter.
\rv{Note that the $D$ operator depends on the gauge field $U_\mu(x)$ and the hopping parameter $\kappa$, and so we use the shorthand $D[U_\mu(x), \kappa]$ to represent the operator defined by a particular gauge and mass.
$D^\dagger D[U_\mu(x), \kappa]$ is defined analogously.
When two operators share the same mass but different gauge fields we omit the hopping parameter from their shorthands; for e.g., $D[U_\mu(x)]$ and $D[\tilde{U}_\mu(x)]$.}

The $\dd$ operator is ill-conditioned for certain regions of couplings such
that it requires effective preconditioning in an iterative solver to accelerate
convergence. Let $M^{-1}$ be the inverse of $\dd$, such that $M^{-1} \dd =
  \mathbb{I}$. We propose to leverage the dependence of the Dirac operator on the
gauge field to produce an operator, $\Gamma$, such that $\Gamma \approx
  M^{-1}$. Specifically, we construct \(\Gamma = \dd[{\tilde{U}_\mu(x)}]\), i.e.,
the \(\dd\) operator generated from new field \(\tilde{U}_\mu(x)\). We note the
focus on producing another set of ``gauge field configuration" shares some
similarity with the framework in \cite{Nagai:2025rok} where an effective gauge
field is used in place of the original field during an intermediate stage in
the hybrid Monte-Carlo process that generates the gauge field configurations.
To this end, the learning objective becomes to obtain $\tilde{U}_\mu(x)$ for a
given $U_\mu(x)$, such that $\dd[{\tilde{U}_\mu(x)}]\dd[{U_\mu(x)}] \approx
  \mathbb{I}$. Moreover, we aim to find an operator such that the mapping is
general for gauge fields $U_{\mu}(x)$ regardless of the size of the underlying
physical system.
Therefore, the mapping to be learned is $\mathcal{G}: \{U_{\mu}(x)\} \mapsto
  \{\tilde{U}_{\mu} (x)\}$, where $\{U_\mu(x)\}$ and $\{\tilde{U}_\mu(x)\}$
represent sets of lattice gauge field configurations, potentially coming from
\emph{arbitrary} action parameters, and their corresponding
inverse-approximating generating configurations, respectively. Our goal is to
use a neural network parameterized by learnable weights to learn the mapping,
$\mathcal{G}$, between the infinite dimensional spaces. In particular,
considering training efficiency and data accessibility, we plan to train the
network on relatively small configurations generated from a specific set of
geometries and couplings and directly apply to other settings\footnote{Training
  the preconditioner on sets of gauge field configurations with different
  geometries and couplings is also possible, but not pursued in this study.}.

\subsection{Operator learners}

We aim to learn from the input lattice gauge fields and produce fields with the
same structure for the preconditioning operator and expect to learn the general
mapping across gauge configurations. Therefore, we leverage operator learning
neural networks that learn the transformation of functions. These networks can
therefore preserve the shape of the input (output has the same shape) while
being able to handle variable input shapes. In particular, we adopt fully
Convolutional Networks~(FCN)~\cite{long_fully_2015}, restricted to learning
from local features, and Fourier Neural Operators~\cite{li_fourier_2021},
equipped to learn from both local and global features, to investigate how
\(\mathcal{G}\) can be approximated.

These networks aim to learn the mapping from instances of the lattice gauge
fields and their inverse approximation generating fields, shown in
(\ref{eqn:map}).

\begin{equation}
  \label{eqn:map}
  \begin{aligned}
    \mathcal{G}_{\theta}(U_{\mu}(x)) = \tilde{U}_\mu(x),
    \quad  U_\mu(x), \tilde{U}_\mu(x) \in U(1)^{X \times T \times d} \subset \mathbb{C}^{X \times T \times d},
  \end{aligned}
\end{equation}

\noindent where $X$ and $T$ are the spatial and temporal lattice extents, and $d=2$ is
the spacetime dimension used in this study. Here $\tilde{U}_\mu(x)$ and the corresponding
\(\kappa_{\tilde{U}_\mu(x)}\) construct a linear function \rv{such that $g(x) =
  \Gamma x = \dd[{\tilde{U}_\mu(x)},\kappa_{\tilde{U}_\mu(x)}]x$}, which is used as the preconditioning
operator in the iterative solver \rv{(see Section~\ref{ssec:probform})}. While it is possible to learn
\(\kappa_{\tilde{U}_\mu(x)}\), for simplicity and consistency, we use the same
hopping parameter as for the original linear operator in this study, i.e.,
\(\kappa_{\tilde{U}_\mu(x)} = \kappa\).

At each layer, both FCN and FNO perform the kernel integral to transform the
input function to another, shown in (\ref{eqn:ker_int}),

\begin{equation}
  \label{eqn:ker_int}
  (\mathcal{K}_{\theta} u)(x) = \int_{y \in \mathcal{D}} K_{\theta}(x,y)u(y)dy  ,
\end{equation}

\noindent where \(\mathcal{ D}\) is the lattice domain, and \(u\) is the input function
(e.g., \(U_{\mu}(x)\) in the first layer of the networks), and \(K\) is a
kernel function parameterized by \(\theta\) that can be learned from data. We
use the composition of the kernel integral with nonlinear activation functions,
$\sigma$, to approximate the underlying true operator as follows

\begin{equation}
  \label{eqn:operator}
  (\mathcal{G}_\theta U_\mu)(x) = \left( \mathcal{K}_\theta^{(N-1)} + \mathcal{B}_\theta^{(N-1)} \right) \circ \sigma \circ \cdots \circ \sigma \circ \left( \mathcal{K}_\theta^{(0)} + \mathcal{ B}_\theta^{(0)} \right)(U_\mu)(x)
\end{equation}

\noindent where \(\mathcal{B}_\theta\) is the learned local bias function (i.e.,
$(\mathcal{B}_\theta u)(x) = u(x) + \mathcal{B}_\theta(x)$ ), and \(N\) denotes
the number of operator learning layers. In particular, FCNs use local spatial
convolutional kernels, taking the form of \((\mathcal{K}_{\theta}u)(x)_{\rm
    FCN} = \sum_{\delta \in S} k_{\theta}(\delta)u(x - \delta) \), \(S\) being the
size of the convolutional kernel (stencil). As a result, FCN is limited to
learning locally from the information dependent on the size of the kernel. On
the other hand, FNO takes the form of
\((\mathcal{K}_{\theta}u)(x)_{\rm FNO} = \sum_{|f|\leq
  m}k_{\theta}(f)u(f)e^{2\pi ifx}\), where \(m\) is the preserved number of
frequency modes sorted from low to high. Therefore, unlike FCN, the FNO
learns from both global and local information. The local nature of interactions
in gauge field theories might suggest the FCN might capture the relevant degrees
of freedom in a preconditioner. However, lattice gauge fields for different
gauge groups exhibits topological features that may be better captured by the
FNO construction which focuses on low Fourier modes\footnote{Since gauge fields
  have gauge redundancies, the connection to Fourier modes is only implicit.}. By
investigating both FNO and FCN approaches, we allow different feature to be
explored. It is likely that structure of the best learned operator will depend
on the particular gauge groups being studied.

\subsection{Loss function}

While minimizing the condition number, or, equivalently, the spectral norm, of
the preconditioned systems is the most straightforward objective for model
training~\cite{cali_neural-network_2023}, reliable computation of the condition
number requires the construction of explicit matrices and singular valued
decomposition, so not suitable for large systems. Moreover, backpropagating the
gradient through the condition number computation can be computationally
expensive and unstable
\cite{hovland2024differentiating,huckelheim2024taxonomy}. Given these
challenges, and motivated by practical considerations in neural network
training, we instead propose to use the differentiable Frobenius norm of the
difference between the preconditioned matrix and the identity to train the
neural networks. Note that since the preconditioner \(\Gamma\) is constrained
to share the same structure as the original operator \(A\), the proposed loss
does not directly optimize it to be the best possible preconditioner within
that structural manifold. Nevertheless, it serves as a practical and effective
proxy. We leave the determination of a more theoretically grounded surrogate
objective for future work. To ensure training efficiency by avoiding explicit
constructions of matrices, we only rely on the linear operators from the Wilson
discretization and utilize random projections to approximate the error in the
inverse via its \(L_2\) norm, shown in (\ref{eqn:loss}).

\begin{equation}
  \mathcal{L}(\theta) = \frac{1}{N \cdot K}\sum_{i=0}^{N-1}\sum_{j=0}^{K-1}\Vert \Gamma_i(\theta)^{-1}D_i^{\dagger}D_iv_{j} -v_{j} \Vert_2,
  \label{eqn:loss}
\end{equation}
where $N$ is the number of samples in \rv{a batch of the training set}, and $K$
is the number of random vectors sampled from an isotropic Gaussian \rv{per
  gradient step during training}. \rv{As Gaussian vectors span the entire vector
  space almost surely, we use them as probes to explore all directions in
  expectation, which allows $\Gamma(\theta)$ to reduce the eigenvalues of
  $D^\dagger D$ broadly.} This number such vectors is treated as a hyperparameter
and fixed through the training. {A brief study on the sensitivity of model
    performance to $K$ is presented in Appendix \ref{app:K_sensitivity}.}

\paragraph{Higher powers of \(\Gamma\).}

\begin{figure}[!htbp]
  \centering
  \includegraphics[width=.8\linewidth]{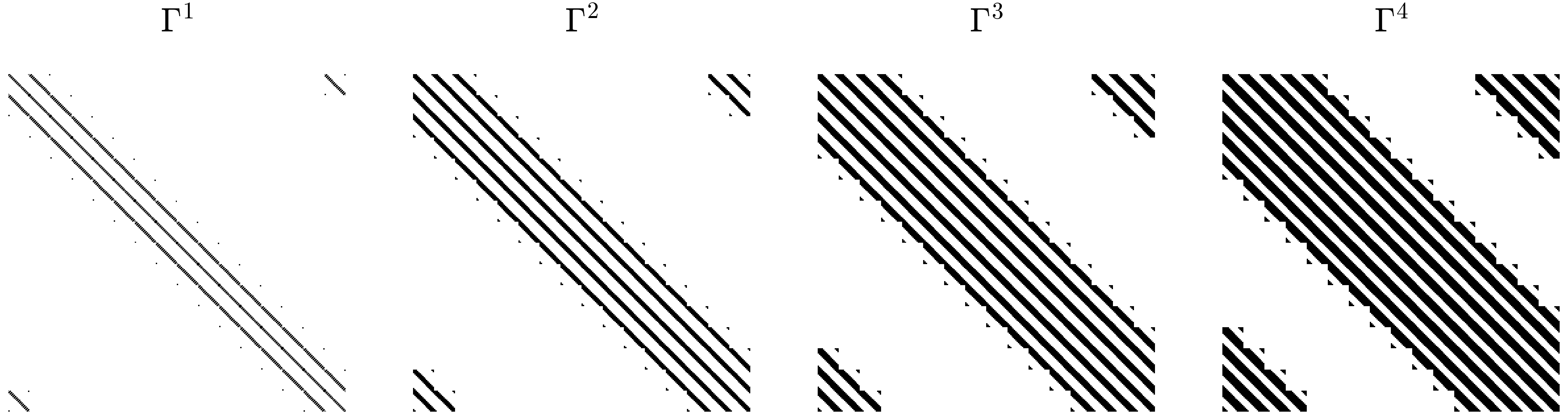}
  \caption{The sparsity pattern of the powers of \(\Gamma\) for \(L=16\). Higher power
    shows denser structure.}
  \label{fig:higher_order_M}
\end{figure}

We rely on the Wilson discretization of the Dirac operator that generates the
original linear system to generate its preconditioner based on the neural
network output \(\tilde{U}_\mu(x)\). This restricts the resulting
preconditioning operator to having the same sparse structure as \(\dd\), which
limits the approximation capacity of the preconditioning operator to its true
inverse. \rv{ Since the true inverse of $\dd$ is fully dense, with a denser
  structure, even though with repeating entries, we hypothesize that
  \((\Gamma)^p\) may be better able to approximate the inverse of \(\dd\),
  leading to additional reduction in number of iterations required in the
  preconditioned CG solver. Note that we know apriori this is a heuristic motivation. A
  rigorous analysis of the preconditioned system is left to future work.}
Therefore, we propose to recursively apply the same discretization to form
higher powers of \(\Gamma\) to obtain \( (\Gamma)^p \), populating the non-zero
structure, shown in Figure~\ref{fig:higher_order_M} for $d=2$. Meanwhile, using
higher powers also introduces additional computation at each solver step. We
discuss the details in Section~\ref{sec:exp}.

\section{Numerical Experiments}\label{sec:exp}

We train models using lattice gauge configurations with various sizes and
couplings (Table~\ref{tab:models}). We generate ensembles using the Hamiltonian
Monte Carlo (HMC) algorithm, with specified parameters. Samples are separated
by 100 HMC trajectories of unit length. Consequently, correlations between
successive samples in each ensemble are determined to be very small. For
action-specific neural preconditioners,
following~\cite{cali_neural-network_2023}, we select the hopping parameter
\(\kappa = 0.276\), corresponding to a mass parameter \(m = -0.188\) which is
close to the critical mass \(m_{\text{crit}} \approx -0.197\) at \(\beta =
2.0\)~\cite{christian2006scaling}. For each model training process, we use 1600
unique configurations for training and validation and another 200
configurations for evaluation. All training uses the same optimizer and
learning rate. We terminate the training after the validation loss has stopped
improving for over 50 consecutive epochs and use the model checkpoint with the
lowest validation loss for evaluation. We first train and evaluate separate
models for each set of action parameters (\(L, \kappa, \beta\)) and then
explore the zero-shot transfer capacity of the framework trained only on one
set of gauge configurations with varied action parameters by directly applying
pretrained neural preconditioners to the CG solver. {We also compare the
    proposed framework against standard preconditioning techniques used for solving
    the Dirac normal equations, including incomplete Cholesky (IChol) and even-odd
    preconditioners. IChol has been shown to be more
    effective~\cite{cali_neural-network_2023}, and our results show that even-odd
    preconditioning has a similar effect as our proposed framework but requires
    additional decomposition and inversion steps. Therefore, we focus our main
    comparisons on IChol and report results for even-odd preconditioners in
    Appendix~\ref{app: evenOdd}.} We emphasize that the learned mapping defined in
(\ref{eqn:map}) can be applied to any lattice geometry, and the operator
learners capture such mapping from a single lattice size. We implement all
model training, evaluation, and preconditioned CG solver in \texttt{JAX} with
double precision, and all tests and timings are generated using an A100 NVIDIA
GPU. The code is available at
\anon{\url{ https://github.com/iamyixuan/neural-precond-schwinger}}. 

\begin{table}[H]
  \centering
  \caption{Overview of ensembles of the two-flavor lattice Schwinger model
    with Wilson fermions used for the numerical study. $L$ is the
    lattice sizes for both space and time dimensions. $\kappa$ and
    $\beta$ are the hopping parameter and gauge coupling used during generation
    of the gauge field configurations. The number of configurations used for
    training, validation and testing is also listed for each model. }
  \label{tab:models}
  \begin{tabular}{c c c c c c}
    \toprule
    Models & $\kappa$ & $\beta$ & \#train & \#val & \#test \\
    \midrule
    $\mathcal{N}_{L8}, \mathcal{N}_{L16}, \mathcal{N}_{L32}, \mathcal{N}_{L64}$
           & 0.276    & 2.0     & 1280    & 320   & 200    \\
    \bottomrule
  \end{tabular}
\end{table}
During training, for all listed models, we fixed the network hyperparameters
and randomly sampled $K=128$ vectors from standard isotropic multivariate
Gaussian distribution to compute and minimize the loss, defined in
(\ref{eqn:loss}). A detailed description of the network architecture, the
choice of hyperparameters, and model training choices can be found in
Appendix~\ref{app:detail}.

\subsection{Action-specific neural preconditioners}

We train the proposed model using lattice gauge fields associated with various
lattice sizes ($L=8, 16, 32, 64$) and investigate the impact of applying the
resulting preconditioners to solve the Wilson-Dirac normal equations. An
overview of the models for the four lattice sizes can be found in
Table~\ref{tab:models}.

\paragraph{Impact on condition number}

\begin{figure}[!htbp]
  \centering
  \includegraphics[width=.6\linewidth]{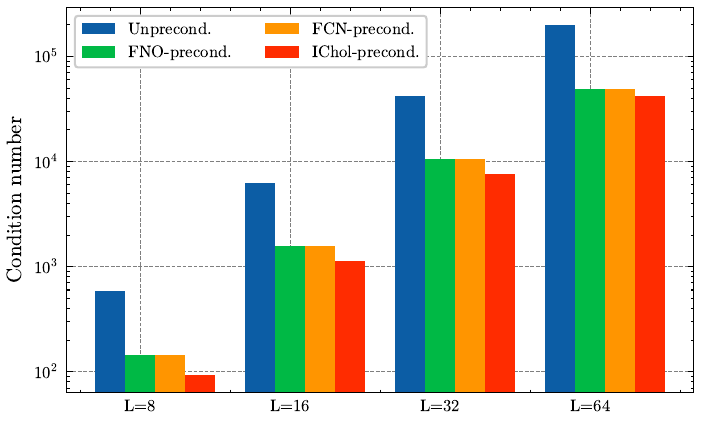}
  \captionof{figure}{Comparison of the average condition numbers among the unpreconditioned systems and
    neural network-preconditioned systems on the testing set with various lattice sizes. Both FNO- and FCN-
    based neural preconditioners significantly reduce the system condition numbers in all cases.} \label{fig:cond_num}
\end{figure}

We apply the preconditioning operator $\Gamma$ obtained from the two types of
trained learners (FNO and FCN) to precondition the linear systems arising from
the listed gauge field configurations. We then compute the condition number of
the unpreconditioned and preconditioned linear operators (\(A\) and
\(\Gamma A\), where $A=\dd[{U_\mu(x)}]$) to evaluate the impact of the resulting
preconditioners. The condition number of a matrix $A$ is defined as $ \kappa
  (A) = {\vert \sigma_{\max} \vert}/{\vert \sigma_{\min}\vert} $ where $
  \sigma_{\max}$ and $\sigma_{\min}$ are the maximum and minimum singular values.
To compute the singular values of the unpreconditioned and preconditioned
systems reliably, we explicitly construct the associated matrices and compute
the condition numbers. Figure~\ref{fig:cond_num} compares the condition numbers
of the linear systems in the testing set before and after applying the FNO- and
FCN-based, as well as IChol preconditioners.
The $\Gamma$ generated from the neural network outputs $\tilde{U}_\mu(x)$
considerably and consistently reduces the condition numbers across all lattice
sizes. Although the IChol preconditioners achieve lower condition numbers, the
construction of neural network--based preconditioners does not require explicit
matrices or their decomposition for new linear systems in the testing set.
Furthermore, the results indicate that the FNO- and FCN-based neural
preconditioners exhibit comparable performance, suggesting they have likely
learned similar mappings.

\paragraph{Accelerating CG solve}

\begin{figure}[!htbp]
  \begin{subfigure}[t]{.49\textwidth}
    \centering
    \caption{}
    \includegraphics[width=\linewidth]{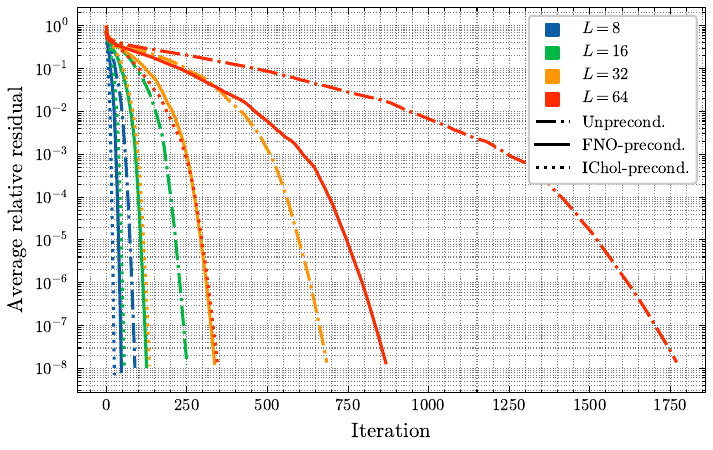}
    \label{fig:cg_solve}
  \end{subfigure}
  \begin{subfigure}[t]{.49\textwidth}
    \centering
    \caption{}\label{fig:cg_time}
    \includegraphics[width=\linewidth]{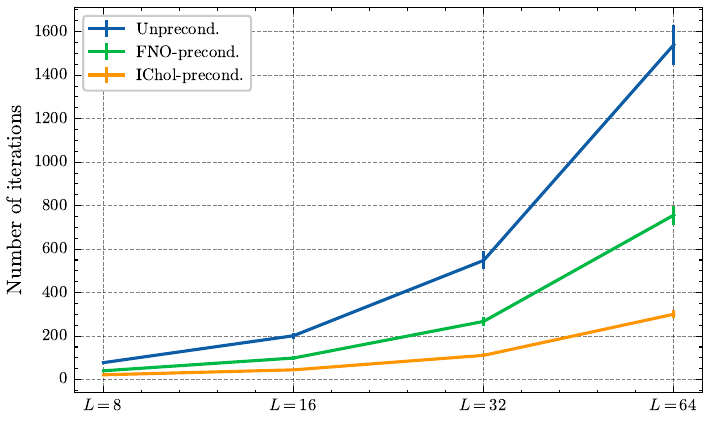}
  \end{subfigure}
  \caption{(a) comparison of the average number of iterations required for a given
    tolerance in the CG solver for the unpreconditoned,
    FNO-preconditioned, and IChol-preconditioned
    linear systems with various lattice sizes. The neural network-preconditioned CG
    solver requires significantly fewer number of iterations than the
    unpreconditioned case. (b) shows the CG iteration counts along with the lattice
    sizes for such systems. While IChol-preconditioners still lead to the fewest
    iterations, the trained FNO-preconditioners show promise of effectively
    accelerating the CG convergence on linear systems with unseen operators in a
    matrix- and solve-free manner. }
\end{figure}

With the proposed framework, the trained neural network--based preconditioners
can be directly integrated into preconditioned CG solvers as linear operators.
We set up the linear systems with random right-hand sides (sampled from an
isotropic Gaussian distribution for both real and imaginary parts and fixed
across models) and solve them using unpreconditioned CG, as well as CG with
IChol, FNO-, and FCN-based preconditioners. With the relative tolerance set to
$10^{-8}$, we compare the average number of iterations and the time required to
reach convergence. Since the FNO- and FCN-based preconditioners exhibit
comparable performance, we report only the results for the FNO-based
preconditioners. Figure~\ref{fig:cg_solve} shows the solver convergence to the
specified tolerance for different lattice sizes. The proposed preconditioning
method effectively reduces the number of iterations required for convergence,
approximately halving the total steps. Moreover, unlike IChol (or even-odd)
preconditioners, the trained FNO-based preconditioners require neither a setup
step (e.g., IChol decomposition) nor an additional triangular solve per CG
iteration; instead, they involve only a forward matrix-vector product, thereby
reducing computational complexity and improving numerical stability. Most
importantly, such neural preconditioners are completely matrix-free through
construction to application, which is desirable especially for large problems
where IChol decomposition may become prohibitively expensive.

\subsection{Volume transfer of trained neural preconditioners}

In this section, we examine the transferability of the trained framework. We
use $\mathcal{N}_{L16}$ in Table~\ref{tab:models} as the base model and apply
it to various lattice gauge configurations. We report the average iteration
counts in the CG solver before and after applying the trained neural
preconditioner on our testing sets. In addition, we investigate the effect of
using models trained with higher powers of $\Gamma$, as described in
Section~\ref{sec:methods}.
\paragraph{Zero-shot performance}

\begin{table}[t]
  \centering
  \caption{Number of iteration in CG solve over the test sets (mean $\pm$ one standard deviation rounded to integers)  for different lattice sizes \((L)\), hopping \(\kappa\), and coupling \(\beta\), comparing
    \textbf{(a)} unpreconditioned CG,
    \textbf{(b)} incomplete-Cholesky preconditioner (IChol),
    \textbf{(c)} neural network-based preconditioner using a pretrained \(\mathcal{N}^{\rm FNO}_{L=16}\),
    and
    \textbf{(d)} neural network-based preconditioner using a pretrained \(\mathcal{N}^{\rm FCN}_{L=16}\).}
  \label{tab:new_test_data}
  \vspace{0.5em}
  \begin{threeparttable}
    \begin{tabular}{ l  c   c  c   c   }
      \toprule
      \multicolumn{1}{c}{\textbf{Configuration}}
       & {\textbf{Unprecond.}}
       & {\textbf{IChol Precond.}}
       & {\textbf{FNO$_{16}$}}
       & {\textbf{FCN$_{16}$}}                                                  \\
      \midrule
      $L=8,\;\kappa=0.276,\;\beta=2.0$
       & $78 \pm 4$                & $22\pm 1$    & $60 \pm 3$   & $40 \pm 2$   \\
      $L=8,\;\kappa=0.276,\;\beta=1.843$
       & $80 \pm 4$                & $23\pm 2$    & $62 \pm 3$   & $42 \pm 3$   \\
      $L=8,\;\kappa=0.260,\;\beta=2.0$
       & $76\pm 2$                 & $21\pm 1$    & $59 \pm 2$   & $40 \pm 1$   \\
      \midrule
      $L=16,\;\kappa=0.276,\;\beta=2.0$
       & $201 \pm 14$              & $44 \pm 4$   & $99 \pm 7$   & $99\pm 7$    \\
      $L=16,\;\kappa=0.276,\;\beta=3.124$
       & $166 \pm 10$              & $33 \pm 2$   & $78 \pm 5$   & $78 \pm 5$   \\
      \midrule
      $L=32,\;\kappa=0.276,\;\beta=2.0$
       & $548\pm 41$               & $111 \pm 9$  & $267 \pm 21$ & $267\pm21$   \\
      $L=32,\;\kappa=0.276,\;\beta=5.555$
       & $260 \pm 19$              & $44 \pm 3$   & $117 \pm 9$  & $117 \pm 9$  \\
      \midrule
      $L=64,\;\kappa=0.276,\;\beta=2.0$
       & $1540 \pm 91$             & $300 \pm 17$ & $719 \pm 47$ & $719 \pm 47$ \\
      \bottomrule
    \end{tabular}
  \end{threeparttable}
\end{table}

\begin{figure}[h]
  \centering
  \begin{subfigure}[t]{.49\textwidth}
    \centering
    \caption{\(L=8\)}\label{fig:zeroshot_L8}
    \includegraphics[width=\linewidth]{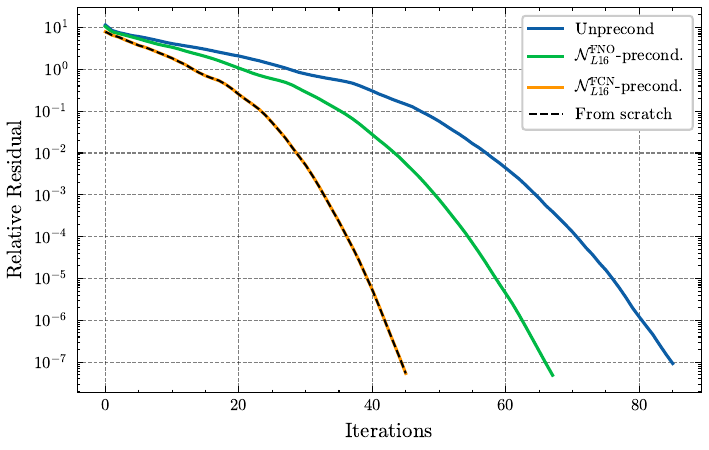}
  \end{subfigure}
  \begin{subfigure}[t]{.49\textwidth}
    \centering
    \caption{\(L=32\)}\label{fig:zeroshot_L32}
    \includegraphics[width=\linewidth]{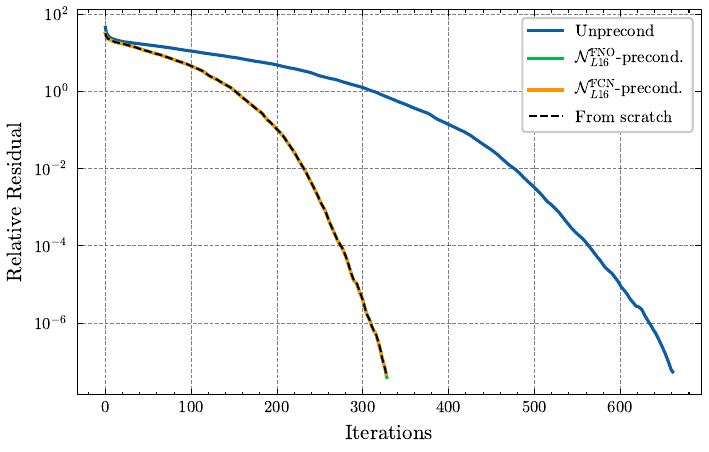}
  \end{subfigure}
  \caption{Zero-shot performance of \(\mathcal{N}_{L16}\) models on gauges configurations of  (a) \(L=8\)  and (b) \(L=32\). The figures show the CG solve residual norm vs. the number of iterations, where
  the neural preconditioner effective reduces the number iterations required for convergence. In particular, the \(\mathcal{N}^{\text{FCN}}_{L16}\) model achieves the same level of reduction as the models trained from scratch using the test cases. The \(\mathcal{N}_{L16}^{\rm FNO}\) retains the
  performance for \(L=32\), and, while still effective, results in an inferior iteration reduction compared with the model trained from scratch on the \(L=8\) case.}
\end{figure}

We expect that the trained networks have learned the general mapping discussed
in Section~\ref{sec:methods} where they are applicable for unseen gauge
configurations. We directly use the trained \(\mathcal{N}_{L16}\) model in the
previous section and obtain preconditioning operators for new \(U_{\mu}(1) \)
configurations with varying lattice sizes, hopping parameters \(\kappa\), and
gauge coupling \(\beta\). Figures~\ref{fig:zeroshot_L8} and
\ref{fig:zeroshot_L32} show the zero-shot performance of \(\mathcal{N}_{L16}\)
(both FNO- and FCN-based) on data with $L=8, $ or $L=32$ in terms of the
convergence in the CG solve. Without any training on the new data with
different lattice size, the model still reduces the number of iteration
required compared to the unpreconditioned case, making it immediately
applicable and effective. In particular, the zero-shot application of the
FCN-based model on gauge configurations with $L=8$ and $L=32$, as well as the
FNO-based model applied to \(L=32\), achieves the same level of acceleration as
models trained specifically on the respective configurations. This demonstrates
the volume transfer capability of the trained model, potentially eliminating
the need of retraining or finetuning. This transferability also allows the
training costs of the NN-preconditioners to be amortized over solutions for
many geometries and parameter sets.
Table~\ref{tab:new_test_data} presents the zero-shot performance of the
\(\mathcal{N}_{L16}\) model on different sets of testing configurations where
\(\kappa\) and \(\beta\) are also changed. Both FNO- and FCN-based
\(\mathcal{N}_{L16}\) lead to significant reduction in the number of iterations
required for convergence of the CG solver. Meanwhile, with larger systems
(\(L\geq32\)), the FNO- and FCN-based \(\mathcal{N}_{L16}\) models have almost
identical performance regarding the number of iterations.

Interestingly, although still managing to decrease the number of iterations,
the \(\mathcal{N}_{L16}^{\text{FNO}}\) model under-performs on \(L=8\) cases
compared to the \(\mathcal{N}_{L16}^{\text{FCN}}\) model and models
specifically trained for these systems. Such performance gap disappears for
larger systems. As described in Section~\ref{sec:methods}, both FNO and FCN are
considered operator learners but with different kernels for conduct kernel
integration. While FNO performs convolution operations in the frequency domain,
essentially having global convolutional kernels, FCN only utilizes spatially
localized convolutional kernels. This mechanism difference forces FCN to rely
on local features of the input function whereas FNO learns from both global and
local features. Therefore, it is likely that the learned general mapping
\(\mathcal{G}_\theta\) in (\ref{eqn:operator}) for these systems depends mostly
on local structures of the gauge configurations, regardless of the differences
in underlying physical systems. In this case,
\(\mathcal{N}_{L16}^{\text{FNO}}\) might have overfit to the global structure
in configurations with \(L=8\), but the seemingly strong dependency of the
general mapping on global structures diminishes for larger lattice sizes. It is
also possible that larger lattice sizes share similar global structures so that
FNOs trained on a single size can still generalize.

\paragraph{Remark.} We note that while the IChol-preconditioners still outperform the
NN-preconditioners in terms of CG iterations, the NN-preconditioners offer the
advantages of avoiding an additional linear solve at each step and eliminating
the setup cost for linear systems with unseen operators. This is particular
useful when only a few right hand sides $b$ to be solved for a given new $A$.
Moreover, thanks to the volume transfer capability, the training time of the
base NN (e.g., 72 minutes for $\mathcal{N}_{L16}^{\rm FCN}$) can be rapidly
amortized when we apply the trained model to new $A$ generated from gauge
fields with different sizes. In contrast, IChol preconditioners must incur the
setup cost (decomposition) for each $A$, which can be computationally intensive
for large problems. Therefore, the choice between the traditional
IChol-preconditioner or the proposed framework depends on the specific
characteristics of the linear systems to be solved. 
\rv{More concretely, using $\mathcal{N}_{L16}^{\rm FCN}$ as the base
model, we quantitatively compare the cost of training,
NN-preconditioned CG solves, and IChol-preconditioned CG solves for
$L=8,16,32,64$. Instead of wall-clock time, which depends heavily on
implementation and hardware, we report FLOP counts as a reproducible
baseline.\footnote{In our current implementation, the NN operations
are carried out with dense kernels (e.g., dense convolutions), while
the FLOP of IChol-preconditioning is estimated as sparse solves.
Optimizing NN operations via quantization/pruning could potentially
decrease the number of new systems needed to amortize the training cost.} The
$\mathcal{N}_{L16}^{\rm FCN}$ model converges in about $1.9\times10^4$
gradient steps with each step costing $2.1\times10^7$ FLOPs, yielding
a one-time training cost of $4.1\times10^{11}$ FLOPs. Using the
per-step FLOP counts and average iteration numbers
(Table~\ref{tab:FLOP}), we compute, for each $L$, the break-even
number $n$ of distinct linear systems $A$ that must be solved for the
NN preconditioner to amortize the training cost compared with IChol.
Using the CG iteration counts from Table~\ref{tab:new_test_data}, we
obtain $n\approx4.5\times10^5, 3.0\times10^4, 1.8\times10^3, 1.1
\times 10^2$ for $L=8,16,32,64$, respectively (see Appendix \ref{app: flop} for
calculation details). Thus, as the lattice size increases, fewer new
systems are needed for the NN-preconditioner to amortize training. In practice,
the number of new systems to be solved exceeds these reported numbers, and it could become
infeasible to form matrices to use IChol preconditioners for larger lattice sizes. This is precisely where our proposed matrix-free framework holds significant advantages.

\begin{table}[t]
\centering
\caption{Operation FLOP count comparison (per step)}
\label{tab:FLOP}
    \begin{tabular}{lcccc}
      \hline
      \textbf{Lattice Sizes} & \textbf{Unprecond CG} & \textbf{NN-precond CG} & \textbf{IChol-precond CG} & \textbf{IChol Setup} \\
      \hline
      $L = 8$                & $1.1 \times 10^{4}$   & $2.0 \times 10^{4}$    & $2.9 \times 10^{4}$       & $1.1\times 10^{6}$           \\
      $L = 16$               & $4.4 \times 10^{4}$   & $7.9 \times 10^{4}$    & $1.2 \times 10^{5}$       & $1.6 \times 10^{7}$           \\
      $L = 32$               & $1.8 \times 10^{5}$   & $3.2 \times 10^{5}$    & $4.7 \times 10^{5}$       & $2.6 \times 10^{8}$             \\
      $L = 64$               & $7.1 \times 10^{5}$   & $1.3 \times 10^{6}$    & $1.9 \times 10^{6}$       & $4.2\times 10^{9}$              \\
      \hline
    \end{tabular}
\end{table}
}

\paragraph{Higher powers of \(\Gamma\)}

\begin{table}[h]
  \centering
  \caption{
    Iteration counts (mean $\pm$ one standard deviation rounded to integers) for the preconditioned CG solver on various lattice sizes using \(\mathcal{N}_{L16}\) trained with \((\Gamma)^p\) for \(p=1, 2, 3, 4\) as the preconditioning operators.
  }
  \label{tab:high_p_iters}
  \resizebox{\textwidth}{!}{
    \begin{tabular}{c c | *{4}{c c}}
      \toprule
               & \textbf{Unprecond.}
               & \multicolumn{2}{c}{\(p=1\)}
               & \multicolumn{2}{c}{\(p=2\)}
               & \multicolumn{2}{c}{\(p=3\)}
               & \multicolumn{2}{c}{\(p=4\)}                                                                                                                       \\

      \textbf{Model}
               & iters
               & FNO                         & FCN
               & FNO                         & FCN
               & FNO                         & FCN
               & FNO                         & FCN                                                                                                                 \\

      \midrule
      \(L=8\)  & $78 \pm 4$                  & $60 \pm 3$   & $40 \pm 2$   & $56 \pm 3$   & $35 \pm 2$  & $62 \pm 4$   & $33 \pm 2$   & $73 \pm 5$  & $32 \pm 2$   \\
      \(L=16\) & $201 \pm 14$                & $99 \pm 7$   & $99 \pm 7$   & $80 \pm 7$   & $86 \pm 7$  & $109\pm 9$   & $81 \pm 6$   & $79 \pm 6$  & $79 \pm 6$   \\
      \(L=32\) & $548 \pm 41$                & $267 \pm 21$ & $267 \pm 21$ & $212 \pm 17$ & $232\pm 18$ & $221 \pm 18$ & $219 \pm 17$ & $221\pm 16$ & $212 \pm 16$ \\
      \bottomrule
    \end{tabular}
  }
\end{table}

Constructing a linear operator by repeatedly applying \(\Gamma\) produces a
less sparse matrix and provides more degrees of freedom in approximating the
inverse. In the $L=16, d=2$ case, at \(p=4\), the corresponding matrix of the
linear operator becomes fully dense. We investigate the effect of using
\((\Gamma)^p\), where \(p=1,2,3,4\), as the preconditioning operators to train
\(\mathcal{N}_{L16}\) and report their impact on the preconditioned CG solver.
Tables~\ref{tab:high_p_iters} show the number of iterations required for the CG
solver to reach convergence on various systems where the trained
\(\mathcal{N}_{L16}\) is applied as the preconditioner. The results show a
further reduction in the number of iterations as \(p\) increases in all cases
with FCN-based neural preconditioners and in most cases with FNO-based ones. In
particular, the FNO-based models at \(p=2\) achieve fewer iterations than at
\(p=3\) and \(p=4\). Meanwhile, the FCN-based models consistently show that
higher powers lead to fewer iterations. This validates our assumption that a
denser matrix, even with repeating entries, is able to better approximate the
inverse of \((\dd)[{U_\mu(x)}]\). Nevertheless, the gain in reducing the number
of iterations can be offset by increased computational complexity per step, as
higher powers of \(\Gamma\) require additional matrix-vector products. This
trade-off might be more pronounced when solving larger systems, which we are
interested in quantifying in future investigations. Given this trade-off, the
choice of the power \(p\) should depend on specific needs and the computational
resources available.

\subsection{The learned mappings}

\begin{figure}[!htbp]
  \begin{subfigure}[t]{.49\textwidth}
    \centering
    \caption{}\label{fig:FNO_scatter}
    \includegraphics[width=.7\linewidth]{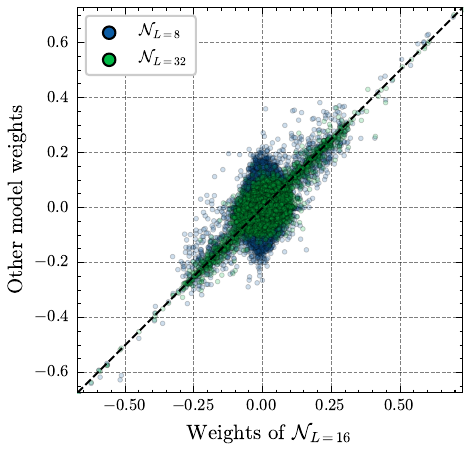}
  \end{subfigure}
  \begin{subfigure}[t]{.49\textwidth}
    \centering
    \caption{}\label{fig:FCN_scatter}
    \includegraphics[width=.7\linewidth]{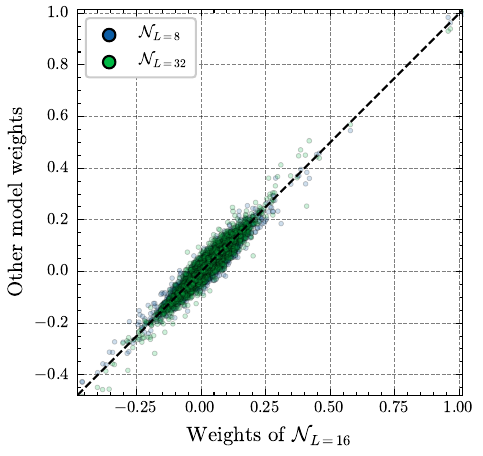}
  \end{subfigure}
  \caption{Scatter plots of model weights learned from different lattice gauge fields. (a) FNO-based; (b) FCN-based.
  }
\end{figure}
The results of the numerical experiments support our conjecture that the
mapping between the input gauge configurations and preconditioner-generating
configurations is general and does not dependent on the specific action
parameters or lattice geometries. Moreover, given the performance discrepancy
between the \(\mathcal{N}_{L16}^{\text{FNO}}\) and
\(\mathcal{N}_{L16}^{\text{FCN}}\) models
for \(L=8\) cases, such mapping might depend only on the local structures of
the gauge fields. To this end, we further examine the trained models in
Table~\ref{tab:models} by comparing model weights to obtain a proxy of
distances among learned mappings. We compare the element-wise weights of models
trained with specific action parameters, as listed in Table~\ref{tab:models},
where models with similar parameters are expected to approximate similar
mappings. Figure~\ref{fig:FNO_scatter} shows the model similarity between the
\(\mathcal{N}_{L16}^{\text{FNO}}\) and \(\mathcal{N}_{L8}^{\text{FNO}}\), and
\(\mathcal{N}_{L32}^{\text{FNO}}\) models. Using $\mathcal{N}_{L16}^{\rm FNO}$
as the reference, the weights of $\mathcal{N}_{L32}^{\rm FNO}$ are more closely
aligned along the diagonal compared to $\mathcal{N}_{L8}^{\rm FNO}$, suggesting
that for FNO models, the learned mapping for $L = 8$ differs from those for $L
  = 16$ and $L = 32$. This observation aligns with the lower performance of
directly applying $\mathcal{N}_{L16}^{\rm FNO}$ to the $L = 8$ case. Meanwhile,
the FCN models trained on different action parameters show high similarity
among the model weights (Figure \ref{fig:FCN_scatter}), indicating these models
have learned similar mappings. Indeed, as described in the previous section,
directly applying $\mathcal{N}_{L16}^{\rm FCN}$ to the $L = 8$ and $L = 32$
cases achieves comparable performance to models specifically trained on those
cases. These results imply the target mapping may dominantly depend on the
local features of the gauge field. While the FNO models may have overfitted the
global structure of the gauge field, which could vary for different problem
sizes, the FCN models, thanks to the architecture restriction, learn the local
dependency that appears to generalize across action parameters.

\begin{figure}
  \centering
  \begin{subfigure}{0.49\textwidth}
    \centering
    \includegraphics[width=\linewidth]{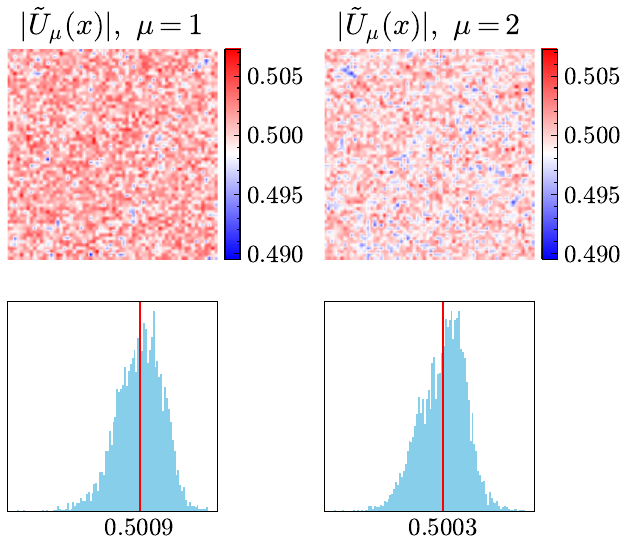}
    \caption{Magnitude of \(\tilde{U}_\mu(x)\)}
  \end{subfigure}
  \begin{subfigure}{0.49\textwidth}
    \centering
    \includegraphics[width=\linewidth]{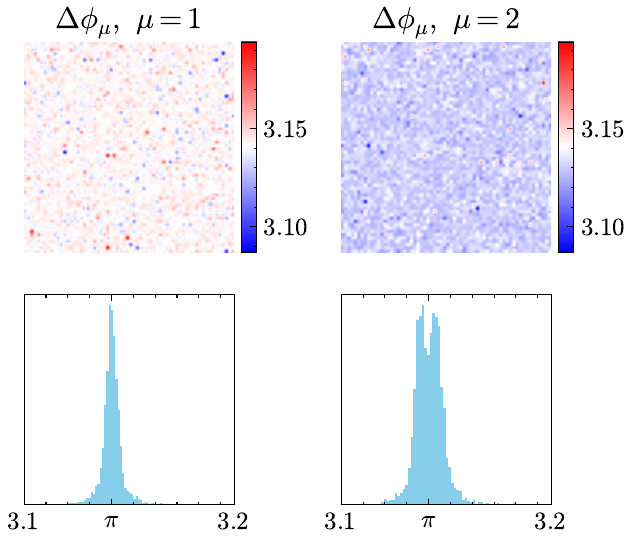}
    \caption{Phase difference \(\Delta \phi  \in [0, 2\pi)\) }
  \end{subfigure}
  \caption{Visualization of an $L = 64$ instance comparing the FNO-predicted
    gauge field $\tilde{U}_\mu(x)$ to the corresponding input configuration
    $U_\mu(x)$ across all lattice sites. (a) shows the magnitude of
    $\tilde{U}_\mu(x)$ (note that $|U_\mu(x)| = 1$ for the input configuration,
    as $U_\mu(x) \in U(1)$); (b) displays the phase difference between
    $\tilde{U}_\mu(x)$ and $U_\mu(x)$. } \label{fig:UtildeVis}
\end{figure}

With the model output $\tilde{U}_\mu(x)$ generating effective preconditioning
operators, we further investigate the relationship between $U_\mu(x)$ and
$\tilde{U}_\mu(x)$. Specifically, we compute and visualize the characteristics
of $\tilde{U}_\mu(x)$ through the changes in phase angles and magnitudes of the
corresponding entries relative to $U_\mu(x)$. For \emph{all} models, we observe
that the average phase difference remains very close to $\pi$, matching up to
the fourth digit, while the magnitudes of \(\tilde{U}_\mu(x)\) is close to
\(\frac{1}{2}|U_{\mu}(x)|\), where $|U_\mu(x)|=1$. For additional details,
Figure \ref{fig:UtildeVis} shows such differences from applying
$\mathcal{N}_{L16}^{\rm FNO}$ to an instance of the $L=64$ case, showing
similar average differences in magnitudes and phase angles. When taking the
signed phase difference, the heat maps and histograms show almost every entry
in $\tilde{U}_\mu(x)$ gets rotated by $\pi$ from $U_\mu(x)$.

These results demonstrate that the learned mappings from various models share
close characteristics, supporting that our proposed framework learns a general
mapping of the form in (\ref{eqn:map}), which may be simple as reducing the
modulus of $U_\mu(x)$ by half and rotating the phases by $\pi$. To test this
simple relationship, we then construct the preconditioner generating
configurations explicitly, following $\tilde{U}^{\rm simple}_\mu(x) =
  -\frac{1}{2} U_\mu(x)$, and use the resulting preconditioning operators to
apply to the CG solve. Given the hopping representation of the Wilson Dirac
operator, $\dd[{U_\mu(x)}]=| 1-2\kappa H | ^2$, where $H=H[U_\mu(x)]$ is the
hopping matrix which depends linearly on $U_\mu(x)$ \cite{Smit:2002ug}, one can
argue that $\dd[{-0.5U_\mu(x)}] = | 1 + \kappa H |^2 \underset{\kappa
    \rightarrow 0}{\simeq} |1+2\kappa H| \simeq |1-2\kappa H|^{-1}\simeq
  (\dd[U_\mu(x)])^{-1}$. Therefore, one might expect $\dd[{\tilde{U}^{\rm
          simple}_\mu(x)}]$ to be an effective preconditioner at small $\kappa$. It is
nevertheless surprising that this relation is approximately learned for
$\kappa$ near criticality. Table \ref{tab:manual} shows that the transformed
gauge configurations are effective at generating preconditioning operators for
the original linear operators, only marginally worse compared to the neural
network-produced \( \tilde{U}_\mu(x)\). However, the slightly inferior
performance of $ \tilde{U}^{\rm simple}_\mu(x)$ constructed from this simple
relationship, suggests that correlations in the changes from $U_\mu(x)$ to
$\tilde{U}_\mu(x)$ across $\mu$ and $x$ are important relative to the learned
$\tilde{U}_\mu(x)$. We leave further investigation in future work.

\begin{table}[t]
  \centering
  \caption{Number of iterations (mean $\pm$ one standard deviation rounded to integers) in the CG solve needed for unpreconditioned and manually constructed \(\tilde{U}\) preconditioned cases. All configurations here share \(\kappa=0.276\) and \(\beta=2.0\).}
  \begin{tabular}{cccc}
    \toprule
                                         & \(L=8\)    & \(L=16\)     & \(L=32\)     \\
    \midrule
    Unprecond.                           & $78 \pm 4$ & $201 \pm 14$ & $548 \pm 41$ \\
    $U^{\rm simple}_\mu(x)$-precond.     & $43 \pm 2$ & $104 \pm 7 $ & $278\pm20$   \\
    \(\mathcal{N}^{\rm FNO}_L\)-precond. & $40 \pm 2$ & $99\pm 7 $   & $267\pm21$   \\
    \bottomrule
  \end{tabular}
  \label{tab:manual}
\end{table}

\section{Conclusion}\label{sec:conclusion}

In this work, we have proposed an operator learning-based neural preconditioner
framework for Wilson-Dirac normal equations in the lattice gauge theory.
\rv{Note that although this study focuses solely on Wilson-discretized Dirac
operators, the proposed framework is applicable to other discretization schemes
whose operators share the same structure as the constructed preconditioners.}
Our method is matrix-free and efficient to train. Once trained, it is
immediately applicable for different lattice geometries and parameter ranges,
achieving the same level of performance as models tailored to specific problems.
Such preconditioners are effective in accelerating the convergence of CG solvers
by reducing the number of iterations required, while maintaining per-step
efficiency through a single matrix-vector step, preventing additional linear
solves. This framework learns a general mapping between the gauge field
configuration and the preconditioner-operator generating field. Therefore, once
trained on certain problems, it leads to effective applications to much larger
systems requiring no further training. \rv{Furthermore, with encouraging results from
applying the proposed framework to the SU(3) gauge group, future work includes a
more extensive study across both SU(2) and SU(3) gauge groups}, as well as
exploring preconditioning operators with structures different from those of the
original systems.

\section*{Acknowledgments} %
\anon{
  We gratefully acknowledge use of the Swing cluster in the Laboratory Computing
  Resource Center at Argonne National Laboratory. The authors thank Gurtej Kanwar
  for constructive discussions and valuable insight. This work is supported by
  the U.S.\ National Science Foundation under Cooperative Agreement PHY-2019786
  (The NSF AI Institute for Artificial Intelligence and Fundamental Interactions,
  \url{http://iaifi.org/}). PES and WD are supported in part by the
  U.S.~Department of Energy, Office of Science, Office of Nuclear Physics, under
  grant Contract Number DE-SC0011090, by the U.S. Department of Energy SciDAC5
  award DE-SC0023116. PES is in addition supported by Early Career Award
  DE-SC0021006, by Simons Foundation grant 994314 (Simons Collaboration on
  Confinement and QCD Strings), and thanks the Institute for Nuclear Theory at
  the University of Washington for its kind hospitality and stimulating research
  environment. This research was supported in part by the INT's U.S. Department
  of Energy grant No. DE-FG02-00ER41132.
  YS, SE, XS, and YL were supported by the U.S. Department of Energy, Office of
  Science, Office of Advanced Scientific Computing Research, Scientific Discovery
  through Advanced Computing (SciDAC) Program, under Contract DE-AC02-06CH11357
  at Argonne National Laboratory (YS and SE), and DE-AC02-05CH11231 at Lawrence
  Berkeley National Laboratory (XS and YL). }

\bibliographystyle{ACM-Reference-Format} \bibliography{references, yixuan_zotero_v2}

\appendix

\section{Model architecture and training details}
\label{app:detail}
We use the same set of hyperparameters to train all models and implement early stopping with patience of 50 epochs. The specifications of the hyperparameters are in Tables~\ref{tab:hparams} and \ref{tab:hparams_FCN}.

\begin{minipage}{.49\textwidth}
    \centering
    \captionof{table}{Hyperparameters of the FNOs}\label{tab:hparams}
    \begin{tabular}{lc}
    \toprule
    Hyperparameter & Value \\
    \midrule
    \# FNO blocks & 4 \\
    \# Fourier modes & 8 \\
    \# Lifting layers per block & 1 \\
    \# Projection layers per block & 1 \\
    Activation & PReLU\\
    Batch size & 128 \\
    Optimizer & Adam \\
    Learning rate & 1e-4 \\
    Early stopping patience & 50 \\
    \bottomrule
    \end{tabular}
\end{minipage}
\hfill
\begin{minipage}{.49\textwidth}
    \centering
    \captionof{table}{Hyperparameters of the FCN}\label{tab:hparams_FCN}
    \begin{tabular}{lc}
    \toprule
    Hyperparameter & Value \\
    \midrule
    \# Layers & 4 \\
    \# Hidden channels & 16 \\
     Kernel size & 3 \\
     Activation & PReLU \\
    Batch size & 128 \\
    Optimizer & Adam \\
    Learning rate & 1e-4 \\
    Early stopping patience & 50 \\
    \bottomrule
    \end{tabular}
\end{minipage}

\begin{figure}
\begin{subfigure}{.48\textwidth}
\centering
\includegraphics[width=\linewidth]{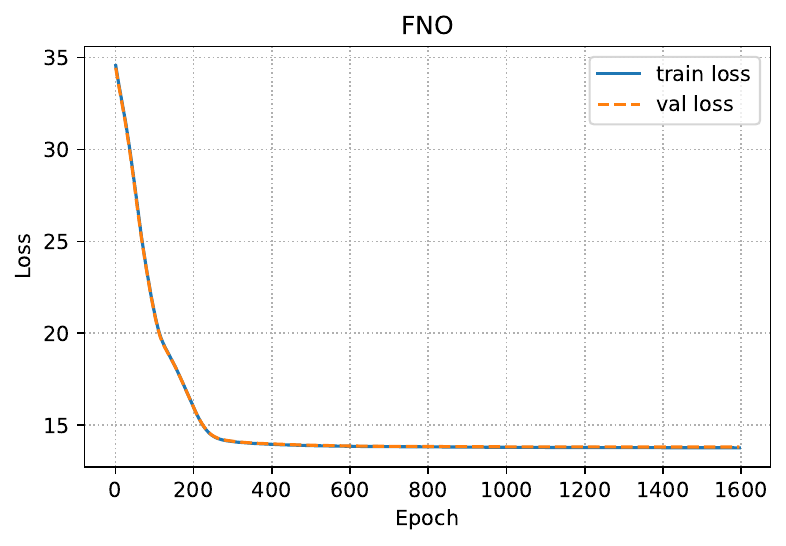}    
\caption{$\mathcal{N}_{L=16}^{\rm FNO}$}
\end{subfigure}
\begin{subfigure}{.48\textwidth}
\centering
\includegraphics[width=\linewidth]{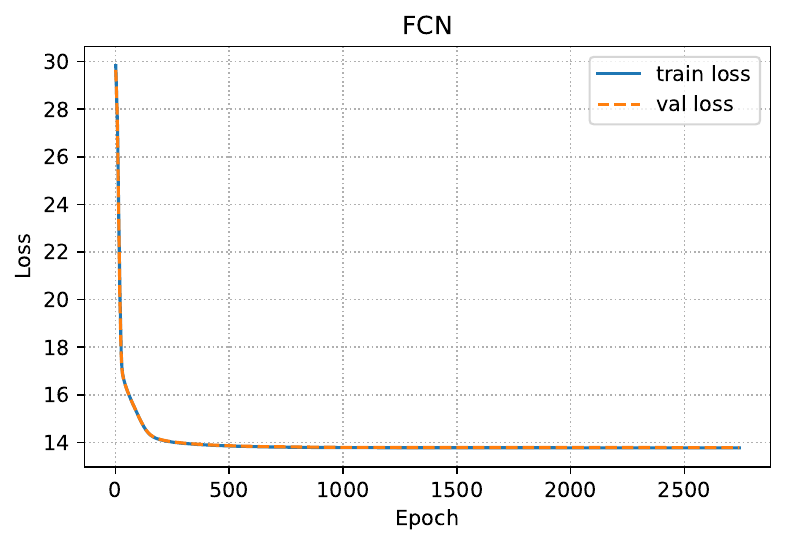}    
\caption{$\mathcal{N}_{L=16}^{\rm FCN}$}
\end{subfigure}
\caption{Training dynamics of the FNO and FCN for $L=16$ cases. Both models converge smoothly, and the validation losses closely track the training losses.}
\label{fig:training-dynamics}
\end{figure}

We implemented the complex neural layers, entire models, and CG solver in
\texttt{JAX} and trained each model in Table~\ref{tab:models} using the same
random seeds and patience for early stopping. Figure \ref{fig:training-dynamics} shows the training dynamics of the two models on $L=16$ cases. The code is available at
\anon{\url{https://github.com/iamyixuan/MatrixPreNet}}\footnote{The repository will be made public
upon acceptance of this work.}.

\rv{
\subsection{Hyperparameter search}

We investigate the impact of various hyperparameters to the model performance
using both the FNO and FCN models. Through Bayesian optimization using
DeepHyper~\cite{balaprakash2018deephyper}, we identify most important sets of
hyperparameters determining the NN-preconditioners' performance. The results are
visualized in Figure~\ref{fig:hpo}.

\begin{figure}
    \centering
    \begin{subfigure}{.48\textwidth}
    \includegraphics[width=\linewidth]{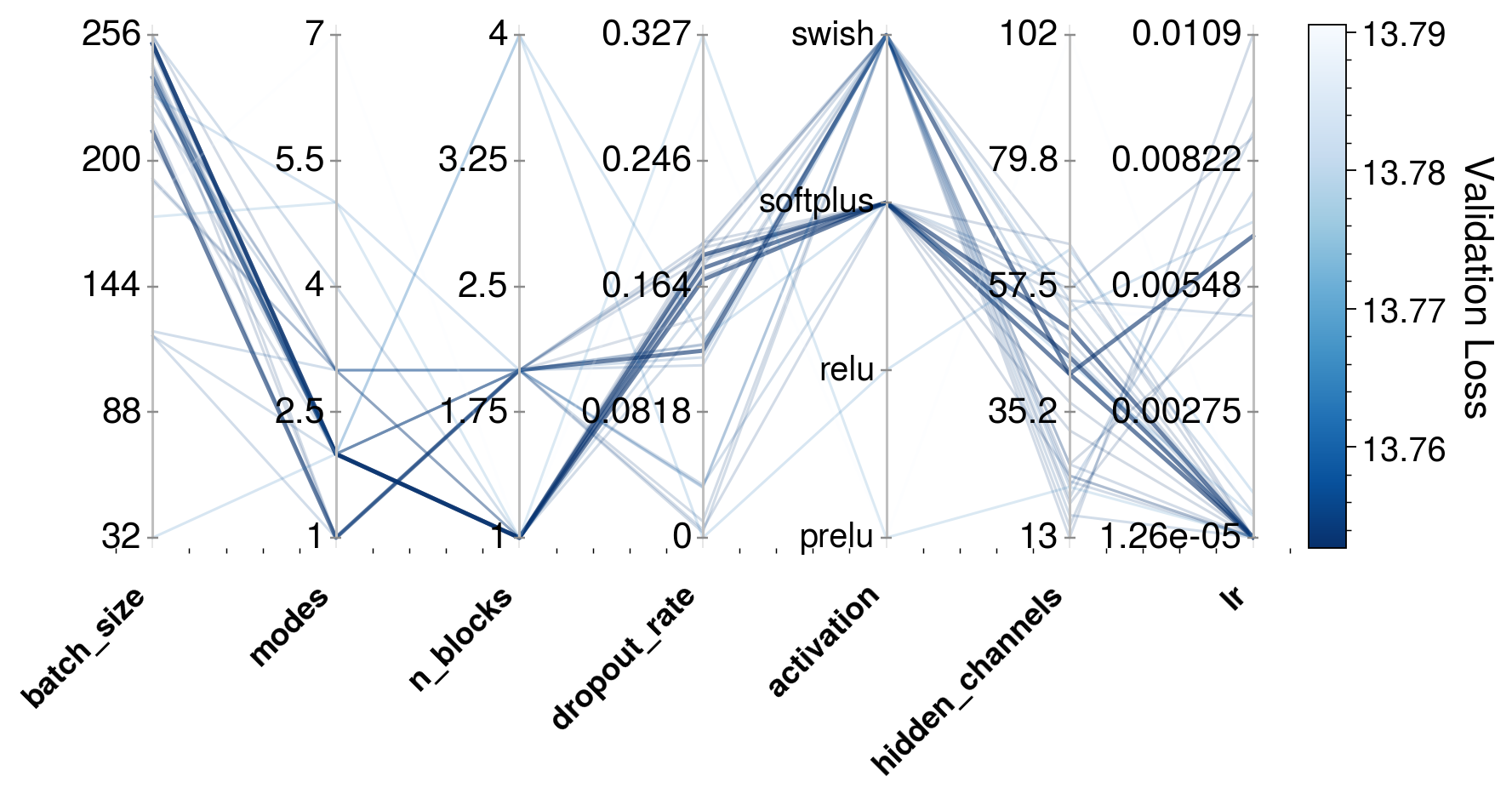}
    \caption{$\mathcal{N}_{L16}^{\rm FNO}$}
    \end{subfigure}
    \begin{subfigure}{.48\textwidth}
    \includegraphics[width=\linewidth]{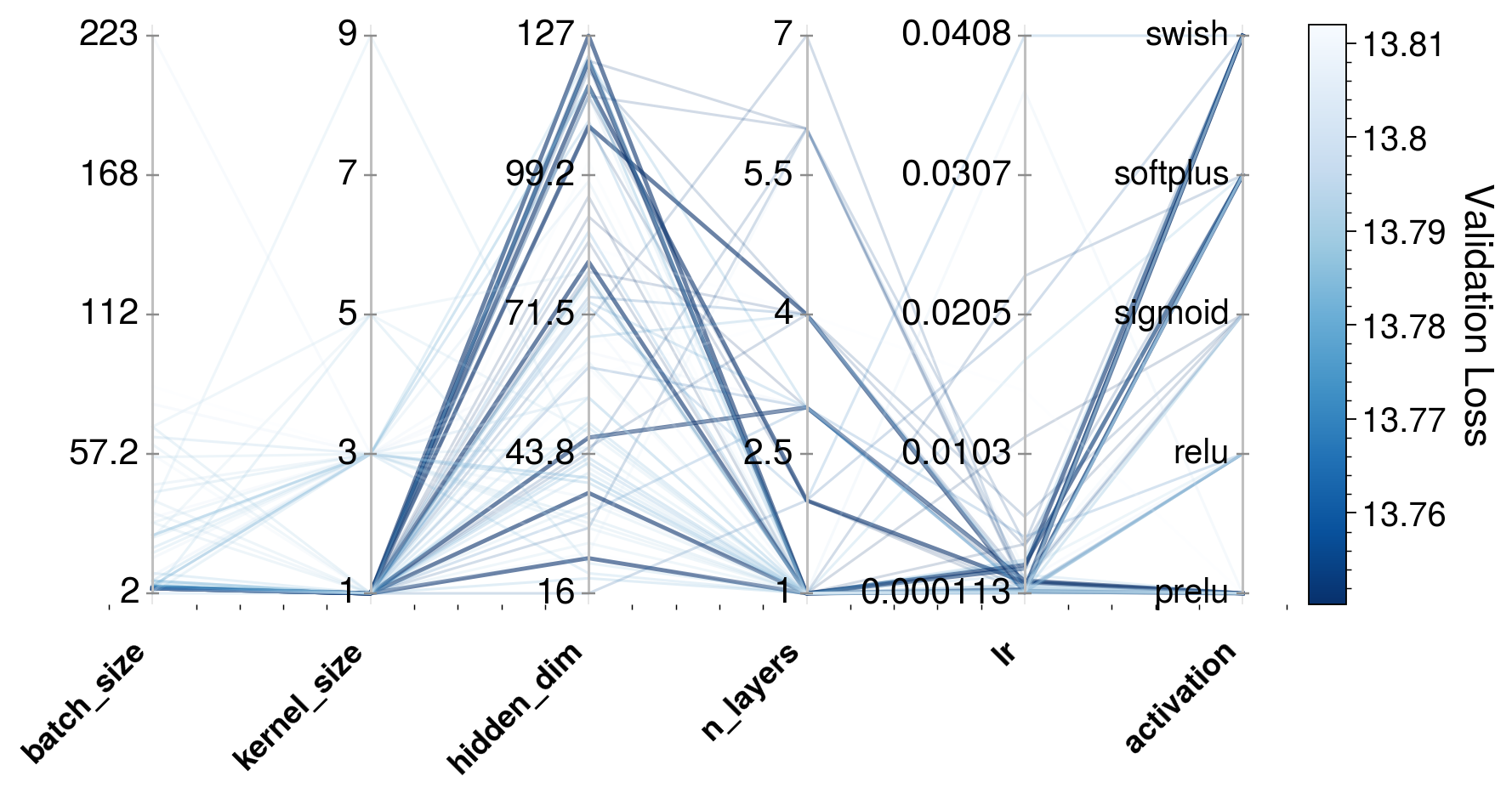}
    \caption{$\mathcal{N}_{L16}^{\rm FCN}$}
    \end{subfigure}
    \caption{Parallel-coordinate plots of hyperparameters for the proposed
    $\mathcal{N}_{L16}^{\rm FNO}$ and $\mathcal{N}_{L16}^{\rm FCN}$ models.
    Using validation loss as the objective, darker blue lines indicate trials
    with lower validation loss.}
    \label{fig:hpo}
\end{figure}

For the $\mathcal{N}_{L16}^{\rm FNO}$ model, a larger batch size, fewer Fourier
modes, fewer FNO blocks, and a smaller learning rate leads to a lower validation
loss. Meanwhile, for the $\mathcal{N}_{L16}^{\rm FCN}$ model, a smaller batch
size and kernel size, higher hidden dimensions, and a smaller learning rate are
corresponding to a lower validation loss. However, in both models, despite the
impact of the various hyperparameters, the variation in validation loss across
hyperparameter configurations is small relative to the absolute loss scale.
These small differences among models are further diminished when applying the
trained models in the PCG solve process. Therefore, given the negligible
sensitivity to hyperparameter choice, we adopt the settings listed above for all
experiments.

}

\section{Impact of the number of random vectors}\label{app:K_sensitivity}

To ensure that the training process remains matrix-free, we rely on projections
of the linear operators onto random vectors, as described in
Section~\ref{sec:methods}. The number of such vectors, denoted by $K$,
influences both the training dynamics and the resulting model performance. Using
$K=128$ as the baseline, we experiment with $K=16, 32, 64, 128, 256$ to train
the model ($\mathcal{N}^{\rm FCN}_{L8}$) and evaluate its performance on the
validation set. For fair comparison, the validation loss is consistently
computed with $K=128$ across all experiments, and the training terminates when
hitting the same early stopping criterion. Using the same early stopping
criterion, Figure \ref{fig:K_sensitivity} shows the offset validation loss from
these models. The curves show that increasing the number of random vectors leads
to lower validation loss, despite the minuscule difference (relative to the absolute loss values). Since using more random vectors slows down the training, with the minimal performance
difference, we use a intermediate value $K=128$ for training.

\begin{minipage}{\textwidth}
    \centering
    \includegraphics[width=.5\linewidth]{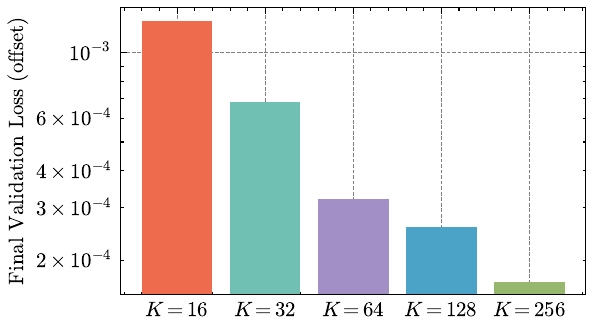}
    \captionof{figure}{Validation loss (offset by 6.869) comparison among models using various number 
    of random vectors for computing the training loss.}
    \label{fig:K_sensitivity}
\end{minipage}    

\rv{
\section{FLOP count calculation}\label{app: flop}
We record the per-step FLOP count for base model training, NN-based PCG solve,
IChol-based PCG solve, and the IChol setup FLOP per new system, to calculate the
break-even number $n$ of new systems such that the NN-based preconditioners show 
advantage.

\begin{table}[t]
\centering
\caption{Operation FLOP count comparison (per step)}
\label{tab:FLOP}
\begin{tabular}{lcccc}
\hline
\textbf{Lattice Sizes} & \textbf{Unprecond CG} & \textbf{NN-precond CG} & \textbf{IChol-precond CG} & \textbf{IChol Setup (sparse)}\\
\hline
$L = 8$  & $1.1 \times 10^{4}$ & $2.0 \times 10^{4}$ & $7.6 \times 10^{4}$ & $8.6 \times 10^{4}$\\
$L = 16$ & $4.4 \times 10^{4}$ & $7.9 \times 10^{4}$ & $1.1 \times 10^{6}$ &$3.2 \times 10^{5}$\\
$L = 32$ & $1.8 \times 10^{5}$ & $3.2 \times 10^{5}$ & $1.7 \times 10^{7}$& $1.3 \times 10^6$\\
$L = 64$ & $7.1 \times 10^{5}$ & $1.3 \times 10^{6}$ & $2.6 \times 10^{8}$ &$5.1\times 10^6$ \\
\hline
\end{tabular}
\end{table}

Denote $m$ as the one-time training cost of the $\mathcal{N}_{L16}^{\rm FCN}$ model, $x_1$ as the NN-based PCG per step cost, $x_2$ the IChol-based PCG per step cost, $s_1$ the average steps required for NN-based PCG solve, $s_2$ the average steps required for IChol-based PCG solve, and $y$ the IChol setup cost per new system. We can compute the minimum number of new systems to amortize the training cost when using NN-based PCG compared to the IChol-based PCG with the following relation
\[
 \begin{aligned}
     m + nx_1s_1 &\leq n(x_2s_2 +y)\\
    n &\geq \frac{m}{x_2s_2 + y - x_1s_1}
 \end{aligned}.
\]
}

\section{Results of applying even-odd preconditioners}\label{app: evenOdd}
Following \cite{cali_neural-network_2023}, we perform even-odd decomposition of the Dirac
operator $D$ as 

\begin{equation}
D = 
\begin{pmatrix}
   D_{ee} & D_{eo}\\
   D_{oe} & D_{oo}
\end{pmatrix}
\end{equation}
which is arranged such that the even sites precede odd sites.
Now, the factorization of $D$ becomes
\begin{equation}
    D = UAL = 
    \begin{pmatrix}
        I & D_{eo}D^{-1}_{oo}\\
        0 & I
    \end{pmatrix}
    \begin{pmatrix}
        \bar{D}_{ee} & 0 \\
        0 & D_{oo}
    \end{pmatrix}
    \begin{pmatrix}
        I & 0 \\
        D_{oo}^{-1}D_{oe} & I
    \end{pmatrix}
\end{equation}
and 
\begin{equation}
U^{-1} = 
    \begin{pmatrix}
     I & -D_{eo}D_{oo}^{-1}\\
        0 & I
    \end{pmatrix}, \quad
L^{-1}=
\begin{pmatrix}
    I & 0 \\
    -D_{oo}^{-1} D_{oe} & I
\end{pmatrix}.
\end{equation}
$\bar{D}_{ee}$ is the Schur complement $\bar{D}_{ee} = D_{ee} -
D_{eo}D^{-1}_{oo}D_{oe}$. After obtaining $U^{-1}$ and $L^{-1}$, instead of
solving the original equation $Dx = b$ (or equivalently $\dd x = D^{\dagger}b$),
we can solve $Ay = c$ (or equivalently $A^{\dagger}Ay = A^{\dagger}c$) where $y
= Lx$ and $c = U^{-1}b$ and then plug in $y$ to obtain the original solution $x
= L^{-1}y$. We treat $A^{\dagger}Ay =A^{\dagger}c$ as the preconditioned 
$\dd x = b$.
Since we are only interested in the condition number and CG convergence rate,
instead of solving the exact systems, we use the random right hand side for both
cases. That is, we are using CG solver to solve $\dd x = b$ and $A^{\dagger}A x = b$
and compare the number of iterations required for convergence.

\begin{minipage}{\textwidth}
    \captionof{table}{Comparison between the unpreconditioned, neural network-preconditoned, and even-odd preconditioned systems in the condition number and CG solver iterations.}    \label{tab:even_odd}
    \resizebox{\textwidth}{!}{
    \begin{tabular}{l*{6}{c}}
    \toprule
         & \multicolumn{3}{c}{\textbf{Condition number (median)}} &\multicolumn{3}{c}{\textbf{CG iterations (max) }}   \\
         & Unprecond. & FNO-precond. & Even-odd precond. & Unprecond. & FNO-precond. & Even-odd precond\\
    \midrule
      $L=8$  & 340.38 & 86.09 &  56.51  & 86 & 46 & 40\\
      $L=16$ & 3708.84 & 901.44& 607.98 & 241 & 121 & 103 \\
      $L=32$ & 30640.93 & 7714.55& 5093.46 & 662 & 329 & 277 \\
    \bottomrule
    \end{tabular}}
\end{minipage}

Table~\ref{tab:even_odd} reports the condition numbers and CG iteration counts
for even--odd preconditioning on the same example problems shown in
Table~\ref{tab:models}. Even-odd preconditioning yields lower condition
numbers than the trained NN‑based preconditioner, and thus requires fewer CG
iterations. However, compared to the NN-preconditioning approach, the reduction
in both the condition number and CG iterations is only \emph{marginally} improved.
Moreover, like IChol, constructing the even-odd decomposed form and forming $U A L$ (and
computing $U^{-1}$ and $L^{-1}$) requires extra work, namely permutation and
inverting $D_{oo}$, which becomes costly for large systems. Moreover, the CG
solve with even-odd preconditioning does not directly produce the full solution;
it still demands a back‑substitution step, adding further overhead. By contrast,
our NN‑preconditioning framework avoids all of these additional decompositions
and solves, and it generalizes across system sizes without retraining.

\end{document}